\documentclass{emulateapj}
\usepackage{natbib,amssymb,graphicx,longtable,epsfig,lscape}
\usepackage{wasysym}
\def\msol{\hbox{$\rm\thinspace M_{\odot}$}} 
\def\sol{\hbox{$\rm\thinspace _{\odot}$}} 
\def\etal{{\it et al.\thinspace}}
\def\eg{{\it e.g.\ }} 

\def\heav{[\Theta(M-M^{+})-\Theta(M-M^{-})]}

\newcommand{\be}{\begin{equation}}
\newcommand{\ba}{\begin{eqnarray}}
\newcommand{\ee}{\end{equation}}
\newcommand{\ea}{\end{eqnarray}}

\begin{document}

\title{Using Spatial Distributions to Constrain Progenitors of Supernovae and Gamma Ray Bursts} 
\author{Cody Raskin\altaffilmark{1}, Evan Scannapieco\altaffilmark{1}, James Rhoads\altaffilmark{1}, Massimo Della Valle\altaffilmark{2,3,4}} 
\altaffiltext{1}{School of Earth and Space Exploration,  Arizona State University, P.O.  Box 871404, Tempe, AZ, 85287-1404.}  
\altaffiltext{2}{INAF- Osservatorio Astronomico di Capodimonte, Salita Moiariello, 16 -80131, Napoli, Italy}
\altaffiltext{3}{European Southern Observatory--Karl Schwarschild Strasse 2- D-85748 Garching bei M\" unchen, Germany}
\altaffiltext{4}{International Centre for Relativistic Astrophysics Network-Piazzale della Repubblica 2, Pescara, Abruzzo, Italy}

\begin{abstract}

We carry out a comprehensive theoretical examination of the relationship between the spatial distribution of optical transients and the properties of their progenitor stars. By constructing analytic models of star-forming galaxies and the evolution of stellar populations within them, we are able to place constraints on candidate progenitors for core-collapse supernovae (SNe), long-duration gamma ray bursts, and supernovae Ia.  In particular we first construct models of spiral galaxies  that reproduce observations of core-collapse SNe, and we use these models to constrain the minimum mass for SNe Ic progenitors to $\approx 25\msol$.  Secondly, we lay out the parameters of a dwarf irregular galaxy model, which we use to show that the progenitors of long-duration gamma-ray bursts are likely to have masses above $\approx 43\msol.$ Finally, we introduce a new method for constraining the time scale associated with SNe Ia and apply it to our spiral galaxy models to show how observations can better be analyzed to discriminate between the leading progenitor models for these objects.

\end{abstract}


\section{Introduction}

Transients are any events that exhibit a brightening, dimming, or otherwise noticeable change within a finite and usually short lifetime. They are one of the most important observable objects in astronomy as they represent a major change of state, which often exerts an enormous impact on its surroundings (\eg McKee \& Ostriker 1977; Dekel \& Silk 1986; Gehrz \etal 1998; Matteucci \& Recchi 2001).  Furthermore, these events inform a great deal on the physics and
 evolution of the progenitor stars that cause them.

Yet, despite their enormous importance, the progenitors of many transient types remain poorly constrained.  One would like to be able to make predictions about these events and watch them unfold in real time, but transients, as their name suggests, are fleeting and fickle - many observed only accidentally and often only partially.  In particular, in the case of supernovae (SNe) and gamma ray bursts (GRBs), observers must rely either on extremely demanding  direct methods or  approximate indirect methods to constrain their progenitor systems.

In the direct approach, one seeks to identify the progenitor star responsible for a particular transient on a case by case basis.  This, in turn, requires one of two things: (i) the occurrence of a rare, nearby event, as in the case of SN1987A (Gilmozzi \etal 1987; Kirshner \etal 1987) in which the surprising properties of the progenitor (\eg Arnett \etal 1989) coupled with neutrino detections (Bionta \etal 1987; Hirata \etal 1987; Mayle \etal 1987) led to important advances in our understanding of stellar physics; or (ii) a laborious search through archival images to extract the one in $\approx 10^{12}$  stars per year that results in a supernova (\eg Mannucci \etal 2005).  Even with this  second, ``brute-force" approach, a good deal of serendipity is still necessary.  Thus, until recently, only one object (1993J; Aldering \etal 1994), had been identified from this approach; the situation improving to a handful of objects with the advent of painstaking searches though pre-explosion Hubble Space Telescope (HST) images (Barth \etal 1996; Van Dyk \etal 2002; Smartt \etal 2004; Maund \etal 2005; Hendry \etal 2006; Li \etal 2006; Gal-Yam \etal 2007).

A second approach relies on indirect constraints and uses the properties of transient host galaxies, or the locations of transient within these  hosts, to derive progenitor properties.  Studies of this type include measurements of the rates of supernovae (\eg Pain \etal 1996; Cappellaro \etal 1999; Tonry \etal  2003; Blanc  \etal 2004; Dahl\' en \etal 2004; Sullivan \etal 2006; Barris \& Tonry 2006; Neill \etal 2006; Botticella \etal 2008;  Dahl\' en \etal 2008) and gamma ray bursts (\eg Daigne \etal 2006; Charay \etal 2007; Liang \etal 2007; Guetta \& Della Valle 2007; Kistler \etal 2008),  constraints from the metal content of galaxy clusters (\eg Renzini 1999; Loewenstein 2001; Scannapieco \etal 2003; Tozzi \etal 2003; Baumgartner \etal 2005) and high-redshift quasars (Barth \etal 2003; Dietrich \etal 2003), studies of the distribution of SNe relative to spiral   arms (Maza \& van den Bergh 1976; Della Valle \& Livio 1994; Bartunov \etal 1994;  McMillan \& Ciardullo 1996; Petrosian \etal 2005), and studies that relate the presence and properties of transients to the environmental properties of their hosts (\eg Oemler \& Tinsley 1979;  Wang \etal 1997; Hamuy \etal 1996; Fruchter \etal 1999; Howell 2001; Le Floc'h 2003;  Christensen \etal 2004; van den Bergh \etal 2005; Ostlin \etal 2008).

Recently, Fruchter \etal (2006, hereafter F06) developed a new such indirect observational method, building on the framework first laid out by Baade (1944) and elaborated on for GRBs by Bloom \etal (2002), and they successfully applied it to differentiate between  the progenitors of long-duration gamma-ray bursts and supernovae.  The method involves observing the spatial locations of GRBs and type-II SNe in their host galaxies and calculating the fraction of the total host galaxy light contained in pixels fainter than these locations.  After several such observations, a pattern emerges, and F06 demonstrated that long-duration GRBs have a higher propensity to cluster in the brighter regions of a galaxy than type II SNe. Kelly \etal (2007) expanded upon this analysis and applied it to type Ia and Ic observations, to find that type-Ic SNe have similar environments to GRBs.

Theoretical work has followed to use the F06 method to put constraints on which stars might contribute to GRB observations (Larsson \etal 2007). In this paper, we show how the F06 method, combined with simple analytic models can be used to derive strong constraints on the progenitors of a wide range of transients.  In particular, we simulate model galaxies that act as hosts for three varieties of transients:  core-collapse SNe (type II's and Ic's), type Ia SNe, and long-duration gamma-ray bursts. Using star-formation rates and stellar evolution profiles, we construct simulated galaxies with all the features of a dynamical system of stellar births and deaths and demonstrate that the transients are primarily determined by their progenitor life-times, and thus progenitors can be constrained by the locations of the transients. By simulating the locations of stellar deaths for massive stars that result in core-collapse SNe and long-duration GRBs, we are able to place constraints on the progenitor masses.  And by simulating the location of low mass stars that result  in white dwarfs and type Ia SNe, we are able to constrain the time scale for these transient events, as well as suggest ways that the F06 method  can be modified to provide further constraints.

The structure of this paper is as follows. In \S 2, we lay out our approach for simulating transients in model spiral galaxies, carry out tests on the resultant data, and apply it to derive mass constraints for core-collapse SN of various types. In \S 3, we lay out the parameters and approach for a simulated dIrr galaxy, establish and test the criteria it must follow, and apply the method to derive mass constraints for long-duration gamma-ray bursts. In \S 4, we return to our spiral galaxy model, employ analytical models for the distributions of type Ia SNe in order to constrain the associated dynamical time-scale, and suggest a modified method to improve these constraints.  Our conclusions are summarized in \S 5.

\section{Core-Collapse SNe}

Type Ib, Ic and type II supernovae are all core-collapse SNe. That is, they are the result of a star whose core has collapsed to form a neutron star or black hole (\eg Woosley \etal 1995). In such objects, the core no longer produces enough energy to counteract its gravitational collapse, and it releases tremendous amounts of neutrinos, eventually resulting in a cataclysmic outflow of the surrounding envelope (\eg Burrows \etal 1995; Mezzacappa 2005). The prevailing model for Ib and Ic SNe is that of an asymptotic giant branch star that has been stripped of most of its envelope by a stellar wind before undergoing core-collapse (Wheeler \& Harkness 1990; Woosley \etal 1995; Yoon \& Langer 2005; Woosley \& Heger 2006). On the other hand type II's retain their gaseous envelopes during the core-collapse and thus feature strong hydrogen absorption lines flanked by P-Cyg absorption troughs.

Kelly \etal (2007) compared observed type II and type Ic counts against the integrated luminosity of the host galaxy in the B-band. In order interpret these observations, we first construct a suitable model for the integrated light of a test galaxy and compute the surface brightness in a coordinate grid $(r,\theta)$ for some time $t$. With this model in hand, we calculate the SNe rate for each pixel in the $(r,\theta)$ grid and compare the result against the integrated light.

Most core-collapse SNe are found in late-type, spiral galaxies of middling to high metallicity, and to this end, we construct a model of a spiral galaxy with solar metallicity, Fe/H = 0.02. Note that Kelly \etal (2007) constrained the effect of the bulge on measuring fractional intensity in the B-band to less than 2\%, and so for our purposes, we need only model the disk component.

\subsection{Approach}

To construct such a model, we start with an expression for the star formation rate that takes the form of a rotating, density wave. In our model, we approximate the density wave as a Dirac delta function with a strength given by some radial profile $\phi(r)$,
\be
	\dot{\Sigma}(r,\theta,t) = 
	\phi(r)
	\sum_{n=0}^\infty\delta(\theta-n\pi-\Omega_p t),\label{eq:sigmadot}
\ee
where $\dot{\Sigma}$ is the star-formation rate per unit area, which, like $\phi(r)$, has units of mass per unit time per unit area. In (\ref{eq:sigmadot}), $\Omega_p$ is the rotational velocity of the star-forming wave, and the sum over delta functions ensures a rotating wave with two arms $\pi$ radians apart.

To calculate the integrated luminosity at some time after the formation of these stars, we employ the population-synthesis code, GALAXEV (Bruzual \& Charlot 2003), which evolves a sample population of stars and outputs time-dependent luminosities, $L_\nu(t)$, in a chosen band, $\nu$. From this, we obtain an expression for the surface brightness of a pixel $(r,\theta)$ at a time $t$,
\begin{eqnarray}
	\nonumber\Sigma_\nu(r,\theta,t)&=&\int_{0}^{t} L_\nu(t-t') \dot{\Sigma}(r,\theta,t')dt'\\
	&=&\phi(r)\sum_{n=0}^\infty L_\nu\left(t-\frac{\theta-n\pi}{\Omega_p}\right)|\Omega_p|^{-1}.\label{eq:sfcbrt}
\end{eqnarray}
The image created by this simple expression represents a snapshot at a time $t$ of the host galaxy and can be compiled for any band.

To compute the SNe rate using the same expression for SFR, we return to GALAXEV, which also calculates the main-sequence cutoff mass, i.e. the stellar mass that leaves the main sequence over time via the Padova 2000 isochrone (Girardi \etal 2000).  A simple inversion then yields the main-sequence turn-off time with mass, $t_{\rm ms}(M)$, which provides a more accurate version of the approximate expression $t_{\rm ms} (M) \approx 10^{10}{\left(\frac{M}{M_{\sol}}\right)}^{-2.5} \textrm{years}$ (\eg Hansen \etal 2004). Since this time interval is approximately 90\% of the total lifetime of a star (Tinsley 1980), this relation places the deaths of large stars near the locations of their formation, while lower mass stars die further away from their birthplaces. 

When stars form in a molecular cloud, large stars are less likely to form than smaller stars, and so,  for a given population, a histogram of stellar masses will be heavily weighted toward the low-mass end (Salpeter 1955; Chabrier 2003). Observationally, one can thus construct  an initial mass function (IMF) which relates the numbers of stars with their mass for a population, and throughout this paper we use the estimate by Chabrier (2003), which is also used by GALAXEV:
\be 
\frac{dN}{d\ln M}\propto\cases{\exp\left[-\frac{\left(\log M-\log
M_c\right)^2}{2\sigma^2}\right] & $M\le\msol$ \cr M^{-1.3} &
$M>\msol$\cr}, 
\ee 
where $\sigma=0.69$, and $M_c = 0.08\msol$. At around $M_c$, the gravitational collapse of the star-forming molecular cloud is not as able to counteract the internal pressure, preventing cloud cores that are smaller than the Jeans mass from forming into stars and flattening the IMF.  

For high mass stars, the supernova rate, $\Psi(t)$ is,
\be	
	-\Psi(t)=\frac{dN}{dt}=1.1{\left[M_{\rm ms}(0.9t)\right]}^{-2.3}\frac{dM_{\rm ms}}{dt},\\
\label{eq:-snr}
\ee
where $M_{\rm ms}(t)$ is the stellar mass corresponding to a main-sequence turn-off time $t$.
We can attenuate this expression with Heaviside functions $\Theta(M-M^{-})$ and $\Theta(M-M^{+})$ to set the lower and upper mass limits for a given SNe type. The final expression is then
\ba	
	\nonumber\Psi(t,M^-,M^+)=1.1{\left[M(0.9t)\right]}^{-2.3}
		\frac{dM}{dt}\\
	\times[\Theta(M-M^{+})-\Theta(M-M^{-})]\label{eq:snr}
\ea

We convolve this expression with the SFR in the same fashion as with the integrated light to produce a per-pixel supernova rate at a time $t$ for the model galaxy.
\ba
	\nonumber\Sigma_{sn}(r,\theta,t,M^-,M^+)=\\
	\frac{1}{\Delta\theta}
	\int_{\theta-\frac{\Delta\theta}{2}}^
	{\theta+\frac{\Delta\theta}{2}}d\theta\int_{0}^{t} \Psi(t-t')\dot{\Sigma}(r,\theta,t')dt'
\label{eq:snbrt}\\
		=\nonumber\phi(r)\sum_{n=0}^\infty 
	\frac{1}{\Delta\theta}\frac{1.1}{1.3}[M(0.9t)^{-1.3}]_
	{\theta-\frac{\Delta\theta}{2}}^{\theta+\frac{\Delta\theta}{2}}\\
	\times \heav 
\ea
Returning to the expression for the radial profile, it is simple enough to begin with an exponential in $r$,
\be	
	\phi(r)=\dot{\Sigma}_0\exp\left(\frac{-r}{r_0}\right),
\label{eq:phi}
\ee
where $r_0$ is the scale radius and $\dot{\Sigma}_0$ sets the star formation amplitude. 

The pattern speed, $\Omega_p$, of a spiral galaxy is dependent on the epicyclic frequency, $\kappa(r)$, and in the frame of the stars, a representative model is
\ba
	\Omega_p(r)&=&\frac{v_0}{r_0}\left(1-
	\frac{1}{\sqrt{2}}\right)-\Omega_*(r),\\
\nonumber\Omega_*(r)&=&\cases{v_0/r & $r>\textrm{1kpc}$ \cr
							v_0/\textrm{1kpc} & $r\leq \textrm{1kpc}.$\cr}
\ea
This velocity will have a singularity at
\be
	r=r_0\left(1-\frac{1}{\sqrt{2}}\right)^{-1} \approx 3.414 r_0,
\ee
where the pattern does not move with respect to the stars. This is called the corotation resonance orbit (Goldreich \& Tremaine 1981; Binney \& Tremaine 1987) and any simulation would have to terminate at radii below this singularity, which fortunately is far enough away from $r_0$ to not affect the outcome.

Finally, the surface brightnesses per pixel in the images obtained from (\ref{eq:sfcbrt}) and (\ref{eq:snbrt}) are sorted into two columns according to the brightest to lowest pixel in the integrated light image. The output of each integral is a 2D linear grid in $(r,\theta)$, the effect being that the pixels are keystone-shaped, and so each pixel must be weighted by $r$ when calculating the fraction of light dimmer than each pixel as is done in the F06-style plot.

\subsection{Tests}

To understand the behavior of this plot, it is necessary to isolate the two mechanisms that place SNe co-spatially in pixels of a certain brightness: the natural dimming due to the radial profile and the location of star formation as opposed to stellar death. Within a dynamical system of star formation that depends only on an angular coordinate such as a grand-design spiral galaxy, constraining observations to an annulus removes the ambiguity introduced by the radial profile. This ambiguity can be described as introducing a second measurement of SNe intensity at a particular brightness that is not due to stellar evolution, but simply to densities of stellar populations. In an annulus, which fixes the model to a single radius, irrespective of the radial profile, the $\theta$ coordinate is analogous to a time coordinate, and features evident along $\theta$ are due only to stellar evolution. Unraveling the annulus, then, is akin to looking through geologic strata, and the locations of the SNe measurements can be seen as the age at which they occurred.

If the model is further constrained to count only those SNe that are due to stars above a threshold mass, a pattern begins to emerge whereby the higher the minimum mass cutoff, the more concentrated the SNe profile in the brightest regions of the galaxy. 

\begin{figure*}[ht]
\centering
\includegraphics[width=8.67cm,height=8.67cm]{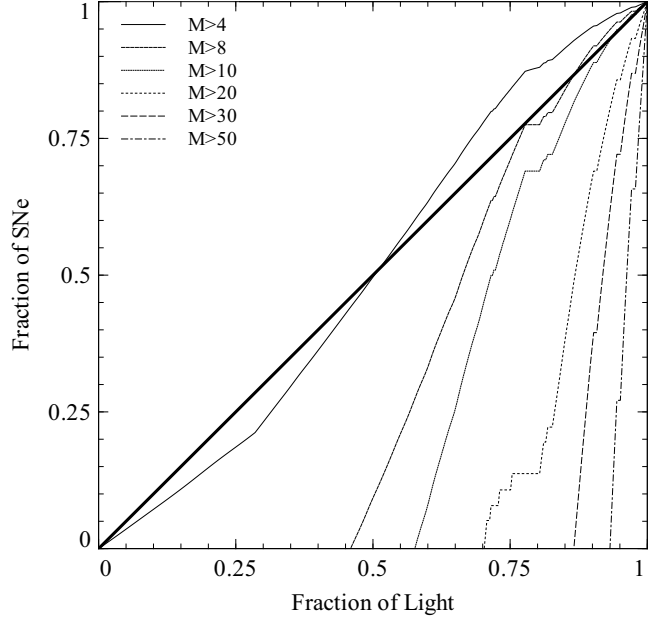}
\caption{A F06-style plot for B-band annuli with variable cutoff masses.
The thick solid line shows a one-to-one correspondence between light and
SNe.  The thin solid lines correspond to SNe with various limits in 
$M^-$, ranging from 4 $M_\odot$ (an unphysical example shown for illustration
purposes) up to 50 $M_\odot$, all with upper limits of 100 $M_\odot$.
}
\label{fig:AnnulusB}
\end{figure*}

One of the features evident in Figure \ref{fig:AnnulusB} is the cutoff brightness for each of the higher mass thresholds where SNe corresponding to stars larger than the minimum mass are no longer observed. This tracks well with stellar age, and the brightness associated with each cutoff are a good indication of the specified minimum progenitor mass. 

\begin{figure*}[ht]
\centering
\includegraphics[width=8.67cm,height=8.67cm]{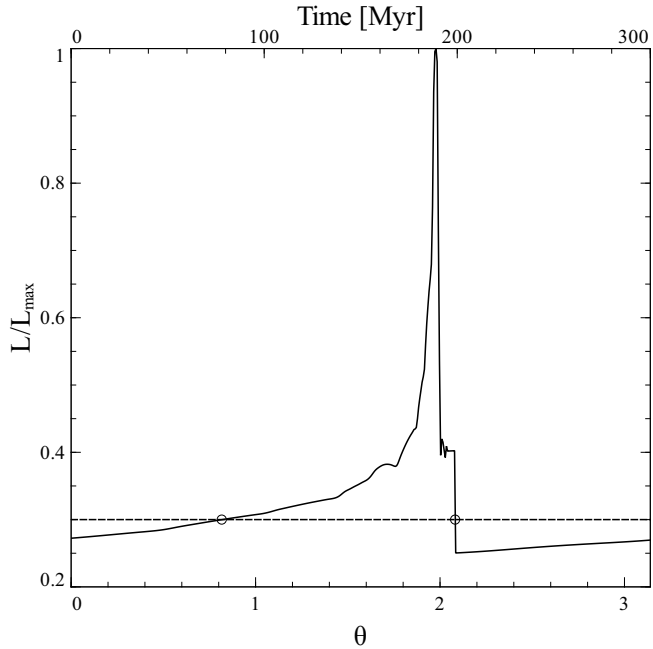}
\caption{A B-band luminosity profile for an annulus. Luminosity is normalized by its maximum.}
\label{fig:CharlotB}
\end{figure*}

Another feature that must be accounted for is the nonlinear climb in the SNe profile with brightness. This is actually due to a second ambiguity in the brightnesses of pixels in the galaxy; for many pixels of a certain brightness \textit{for each radius}, there are two corresponding ages which can be on the order of 10-100 Myr apart. The effect of this is to introduce gaps in the comparison of SNe intensities to the integrated luminosity of the host galaxy. When plotted in a F06 fashion, these gaps result in non-linearities in the SN curve. The effect is small but non-zero and must also contribute to the integrated F06 profile. Figure \ref{fig:CharlotB} demonstrates this ambiguity in the B-band.

Restoring the radial profile essentially smears the behavior shown in Figure \ref{fig:AnnulusB} across the dimmer regions of the galaxy as it reintroduces the ambiguity of SNe measurements at dimmer locations coming from stellar evolution and, critically, from regions of the galaxy at large radius that simply contain fewer stars.

\begin{figure*}[ht]
\centering
\includegraphics[width=8.67cm,height=8.67cm]{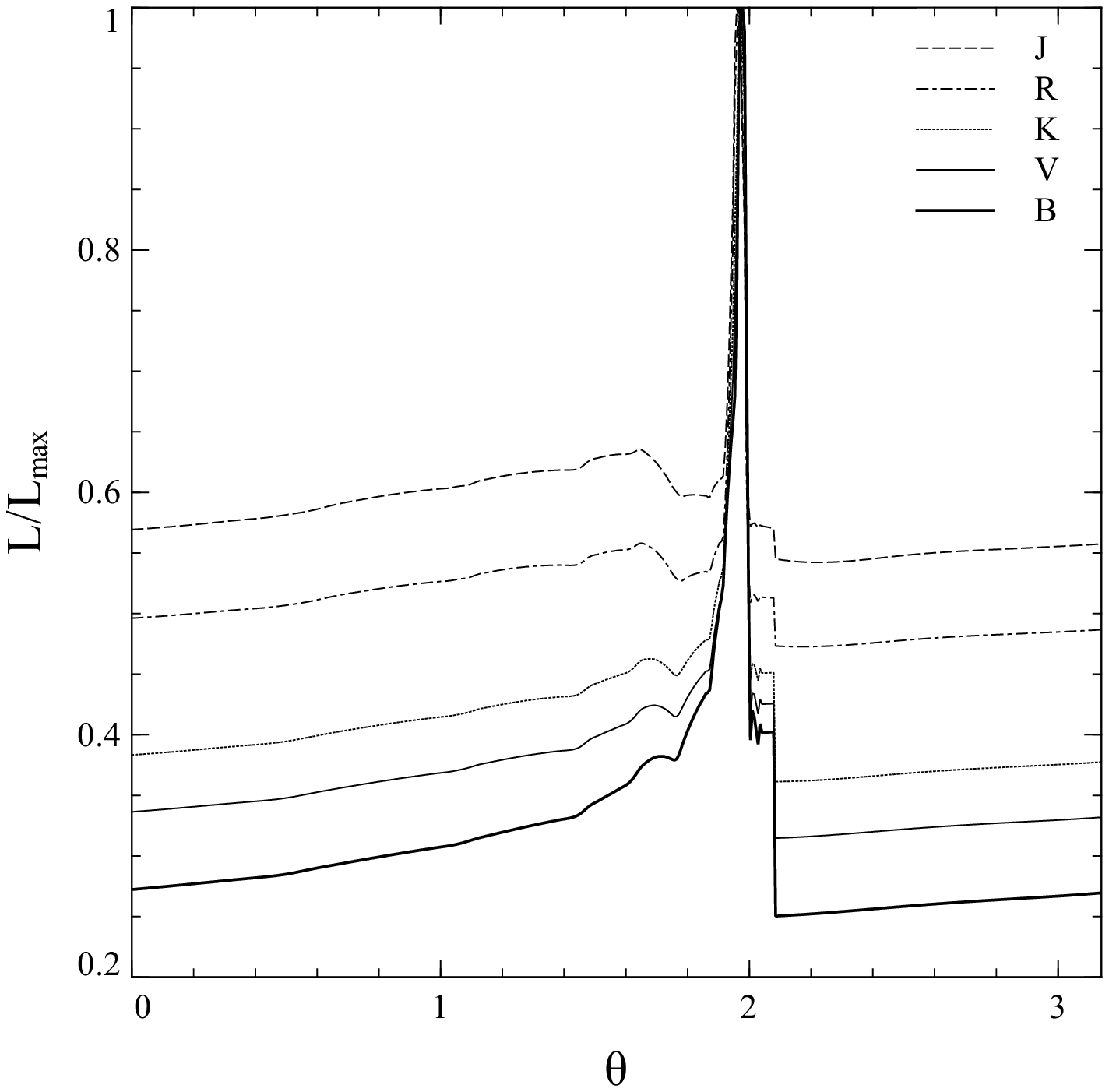}
\caption{Luminosity profiles for an annulus in several bands, normalized by their maximum luminosities. Shorter-wavelength bands exhibit greater contrast from peak to trough.}
\label{fig:BandAnnuli}
\end{figure*}

Finally, a behavior that is most evident when constraining observations to an annulus is the propensity for the mass-brightness relations to break down at longer wavelengths. At near-infrared wavelengths, brightness is less strongly associated with star formation, and therefore the relationship between a star's mass and its location within a galaxy is not as clearly evident. In these idealized models, we neglect the effects of dust, which will complicate the results to the extent that obscuration is biased to regions of ongoing star formation.  
The behavior of a sample of stars over time for several bands is plotted in Figure \ref{fig:BandAnnuli}.

\subsection{Results}
Our first simulation, using solar metalicities, counts SNe from all stars above $8M_{\sol}$, which corresponds roughly to the minimum mass for a core-collapse, type II SN. The total number of SNe per pixel can be thought of as a supernova intensity, with every SN counting equally toward the total intensity, i.e. the intrinsic luminosity of a particular supernova is not a factor. 

\begin{figure*}[ht]
\centering
\includegraphics[width=8.67cm,height=8.67cm]{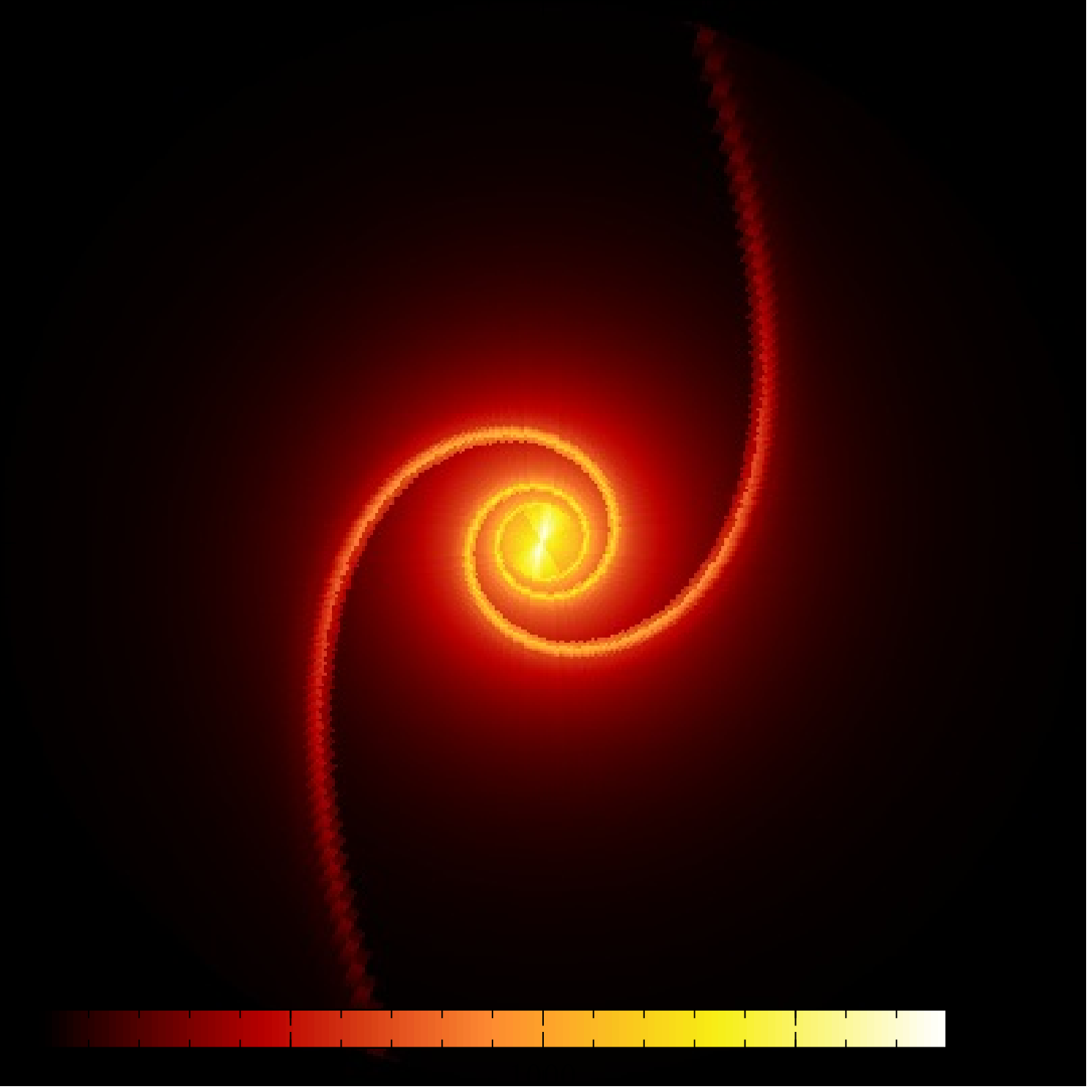}
\caption{B-band luminosity of a model spiral galaxy with $Z=Z_{\sol}=0.02$.}
\label{fig:SpiralGalaxy}
\end{figure*}

\begin{figure*}[ht]
\centering
\includegraphics[width=8.67cm,height=8.67cm]{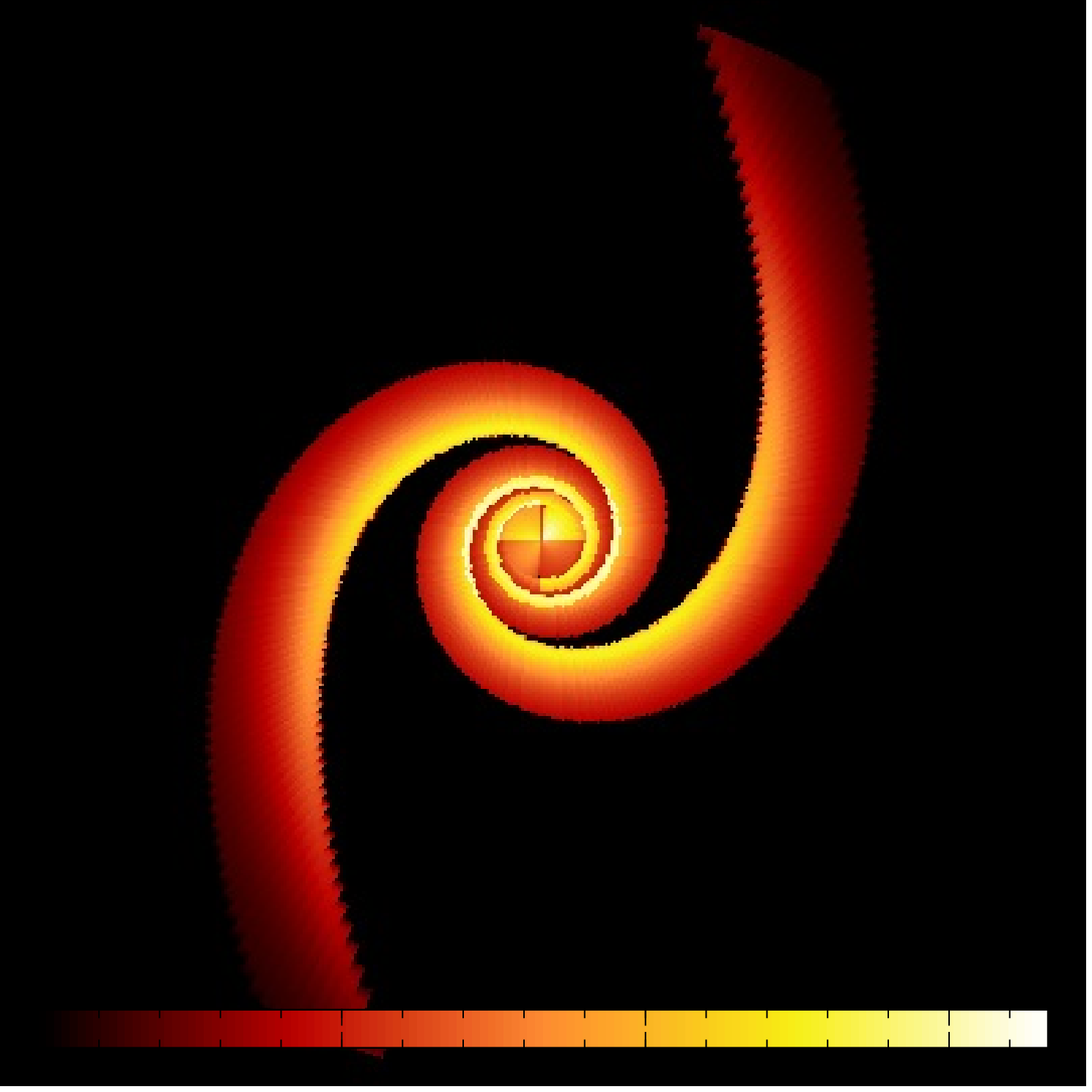}
\caption{SNe intensity for $M>8M_{\sol}$, corresponding to the B-band light distribution of the model spiral galaxy.}
\label{fig:8Mcut}
\end{figure*}

When comparing the B-band, integrated light profile of the simulated galaxy to a SN intensity plot for the same parameters of galaxy size and metallicity, as shown in Figures \ref{fig:SpiralGalaxy} and \ref{fig:8Mcut}, it is immediately evident that the final locations of all stars larger than $8 M_{\sol}$ produce a similar profile to that of the integrated B-band light of the entire galaxy. The majority of the galaxy's B-band intensity is emitted by the large stars that have relatively short lifetimes, approximately 35Myr for $8M_{\sol}$, as compared to a typical galactic rotation time on the order of 300Myr. This places their deaths near to their birthplaces and hence to most of the B-band emission that occurred throughout their lifetimes.

In both Figs.\ \ref{fig:SpiralGalaxy} and \ref{fig:8Mcut}, the intensity peaks at or near the crest of the star-forming wave, where large stars are created, burn brightly in the B-band, and die rapidly. As we adjust our simulation to the redder end of the spectrum, this prominent feature which gives spiral galaxies their arms becomes less and less apparent, until in the K-band, the galaxy looks more like a featureless, exponential disk than a grand-design spiral. This also follows observations that spiral arms become less pronounced
as one moves towards longer wavelengths (Zaritsky \etal 1993; Rix \& Rieke 1993).

\begin{figure*}[ht]
\centering
\includegraphics[width=8.67cm,height=8.67cm]{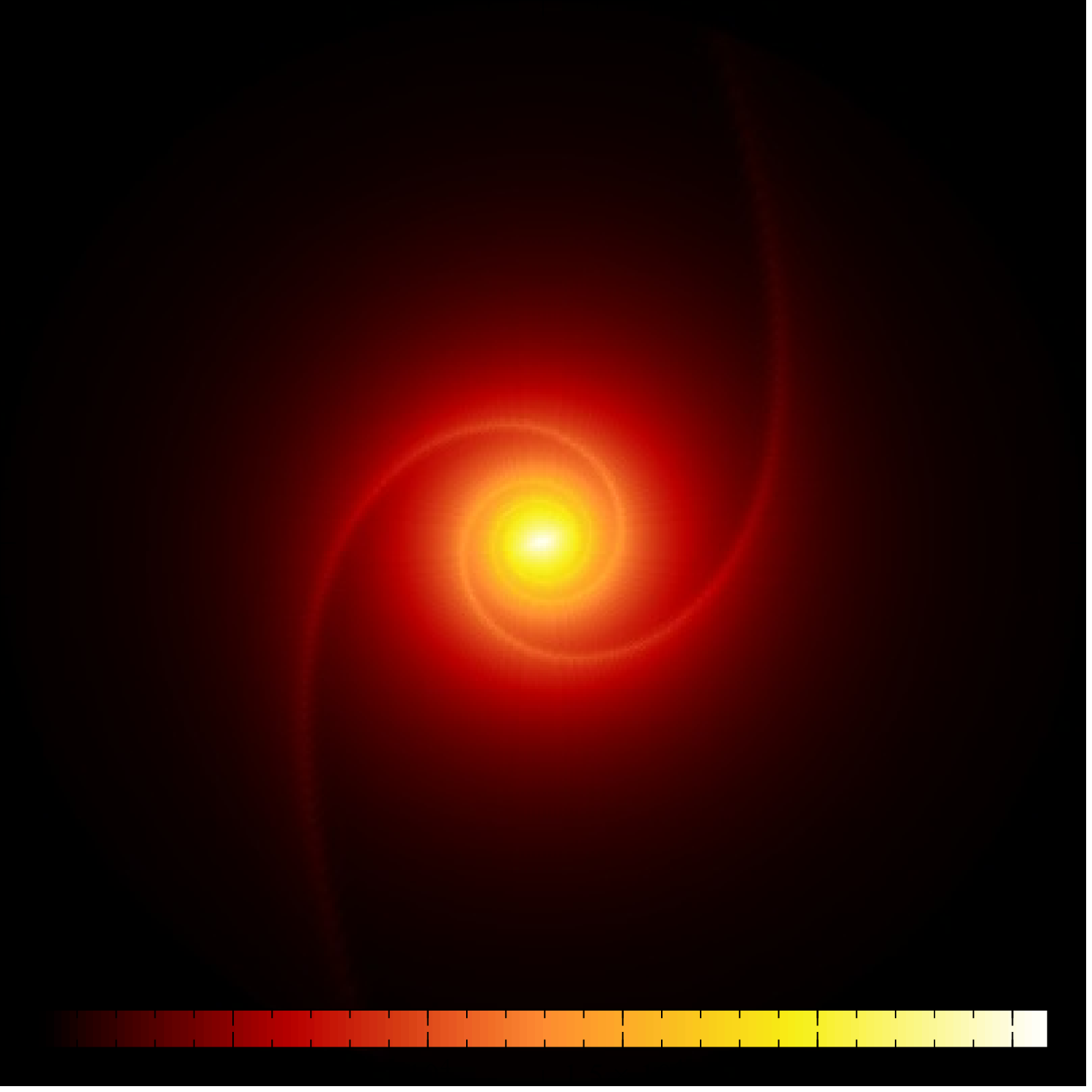}
\caption{K-band luminosity of a model spiral galaxy illustrating weak arms in the low-energy spectrum.}
\label{fig:SpiralKBand}
\end{figure*} 

\begin{figure*}[ht]
\centering
\includegraphics[width=8.67cm,height=8.67cm]{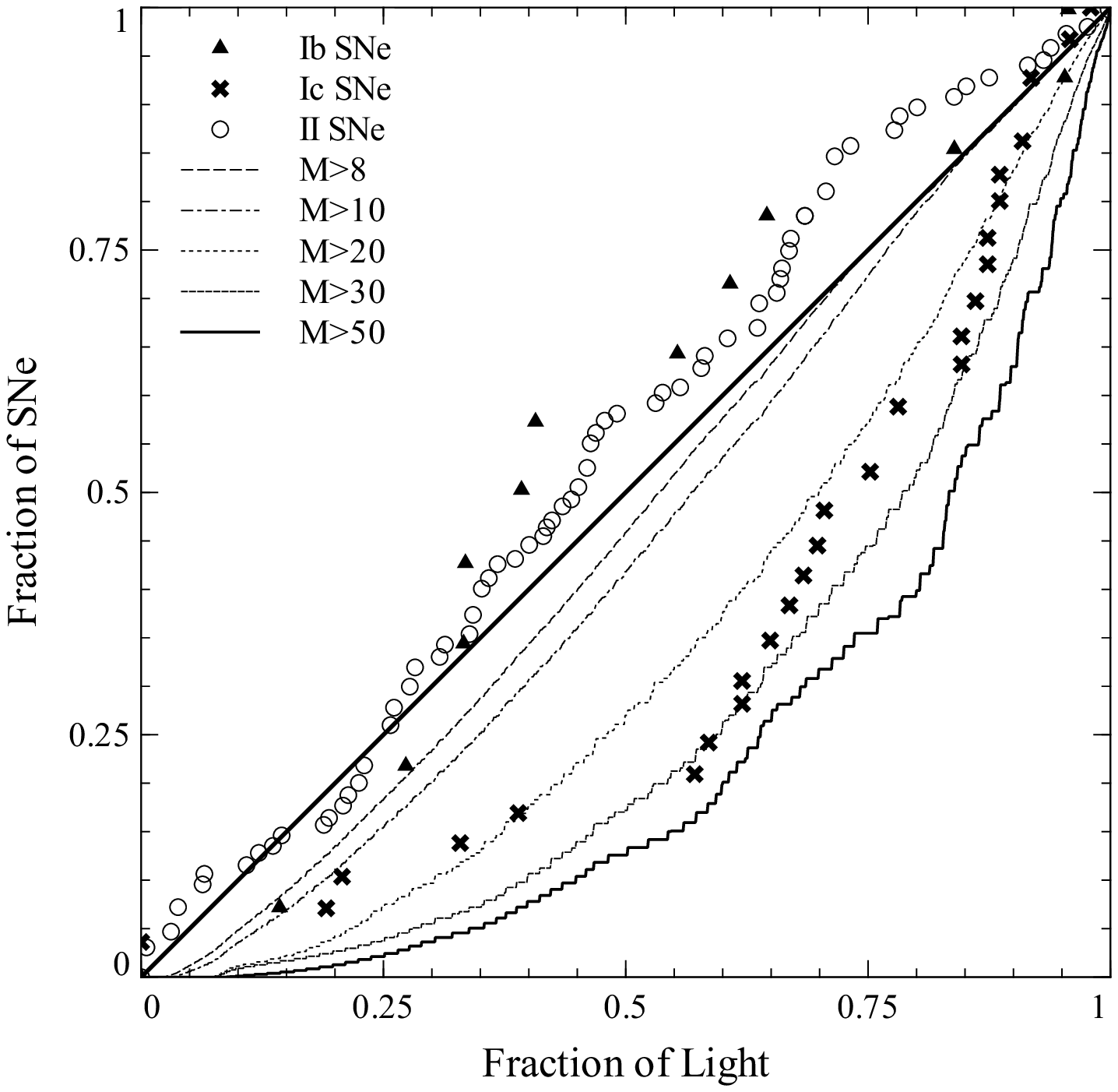}
\caption{F06-style plot for core-collapse SNe in spiral model. Observed type Ib, Ic and II SNe 
from Kelley \etal 2007 are plotted for comparison. The diagonal line follows a 1:1 relationship between fractions of SNe and fractions of total light.}
\label{fig:FruchterIc}
\end{figure*}

Returning to the B-band simulation for $M \geq 8M_{\sol}$, an F06-style plot of the pixel intensities (Figure \ref{fig:FruchterIc}) shows a correlation between type II SNe intensity and B-band light which was demonstrated observationally in F06. If we adjust $M^-$ in (\ref{eq:snr}) to estimate the type Ic minimum progenitor mass as $25 M_{\sol}$, the arm features in the SN intensity plot become much more constricted as compared to the B-band light.

\begin{figure*}[ht]
\centering
\includegraphics[width=8.67cm,height=8.67cm]{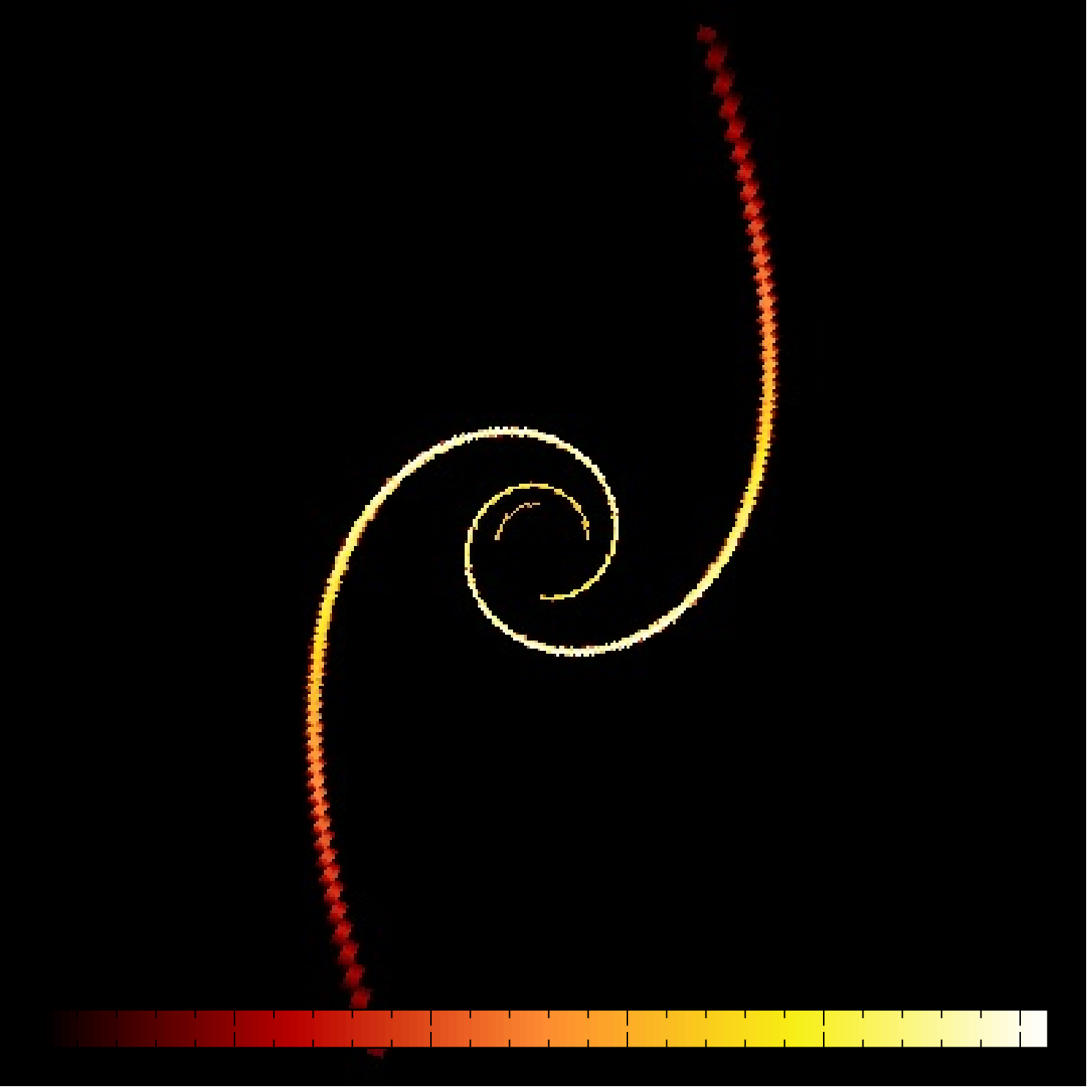}
\caption{SNe intensity for $M>25M_{\sol}$.}
\label{fig:25Mcut}
\end{figure*} 

A plot of these two profiles yields a very similar behavior to that observed in F06, namely that the ordinary type II core-collapse SNe show a nearly 1:1 relationship with the B-band light. However, our simulation also demonstrates that observed type Ic SNe show a strong correlation to the spatial distributions of stars with masses above $\approx 25\msol$. This supports the idea that type Ic SNe are the transient events associated with the deaths of large stars.

\begin{figure*}[ht]
\centering
\includegraphics[width=8.67cm,height=8.67cm]{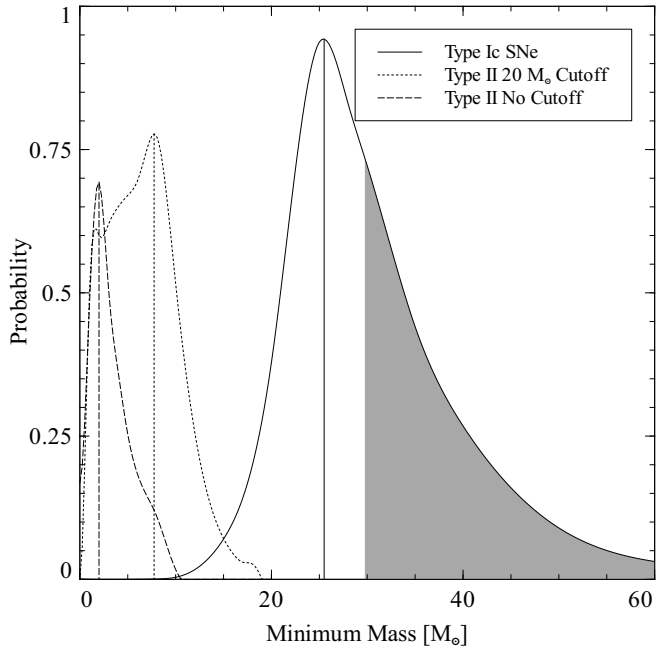}
\caption{K-S test for type Ic and type II SN minimum progenitor mass. 
A better fit to the data is achieved is we impose a
$20\msol$ upper cutoff on the type II simulations. 
In the Ic case, no upper mass cutoff is imposed, but
the gray area indicates disallowed minimum masses above which there would be too few
Ic SNe to be consistent with the observed rates.}
\label{fig:ksic}
\end{figure*}

A K-S test of the resultant data, comparing differing mass cutoffs to the observations described in Kelly \etal (2007) and shown in Figure \ref{fig:ksic}, confirms that $25\msol$ is the most probable candidate for the minimum mass responsible for type Ic SNe. 
Sullivan \etal (2006) suggest that 15\% of all core-collapse SNe take the form of type Ic's, and so, if

\be
\frac{\int_{M^-}^{100}{M^{-\alpha}}dM}
	{\int_{8}^{100}{M^{-\alpha}}dM}=0.15,
\ee

then $M^-\le 30\msol$ is the largest minimum progenitor mass for type Ic's that could yield a 15\% SNe rate, given the Kroupa IMF (Kroupa 2001). The gray area in Figure \ref{fig:ksic} indicates disallowed minimum masses for type Ic's based on this analysis.

\begin{figure*}[ht]
\centering
\includegraphics[width=13cm,height=6cm]{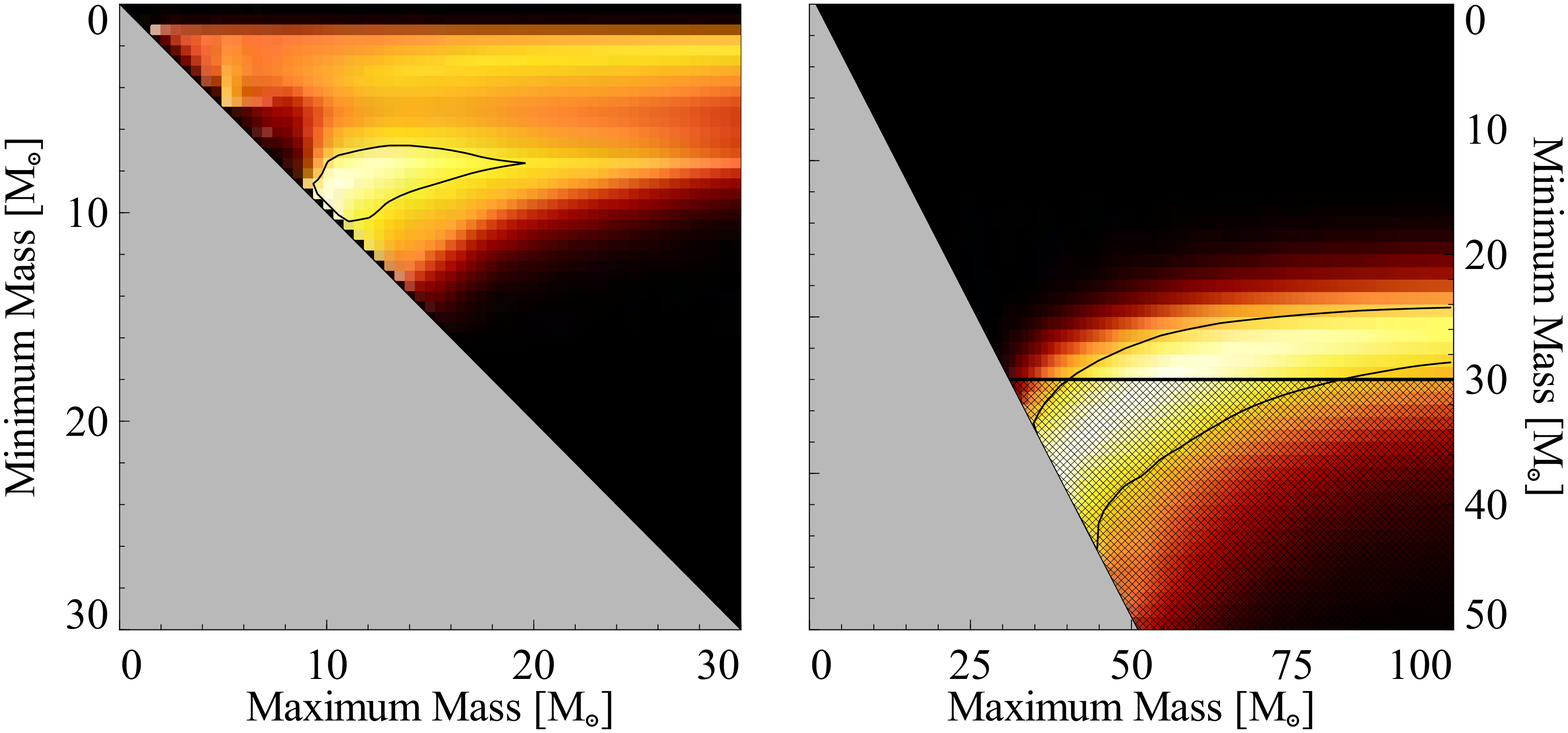}
\caption{2D K-S plots for type II (left panel) and Ic SNe (right panel) with contours at the 85\% confidence level. While the left plots favor a finite upper mass limit for Type-II SNe, no upper mass cutoff is required to match observations of type Ic SNe. The shaded region below $30\msol$ in the right panel corresponds to disallowed minimum type Ic progenitor masses.}
\label{fig:ks2dic}
\end{figure*}

In the type II case, on the other hand, the best fit is obtained when one imposes both a lower mass cutoff and an upper mass cutoff of  $\approx 20\msol$ 
as shown in the type II curves in Figure \ref{fig:ksic}.
This fact is borne out in Figure \ref{fig:ks2dic}, which plots K-S probabilities for various ranges of masses and demonstrates a greater than 85\% probability of a maximum mass of $20\msol$ for type II SNe. Figure \ref{fig:ks2dic} also demonstrates the lack of any compelling upper mass cutoff for type Ic SNe, and so a similar maximum mass is not imposed for type Ic's.

\subsection{Observational Requirements}

Two important observational requirements exist for the analysis laid out here. The first is a minimum image resolution that provides enough data for performing an accurate analysis, and the second is a maximum inclination that allows one to accurately distinguish between different regions within the galaxies.

\begin{figure*}[ht]
\centering
\includegraphics[width=14cm,height=7.25cm]{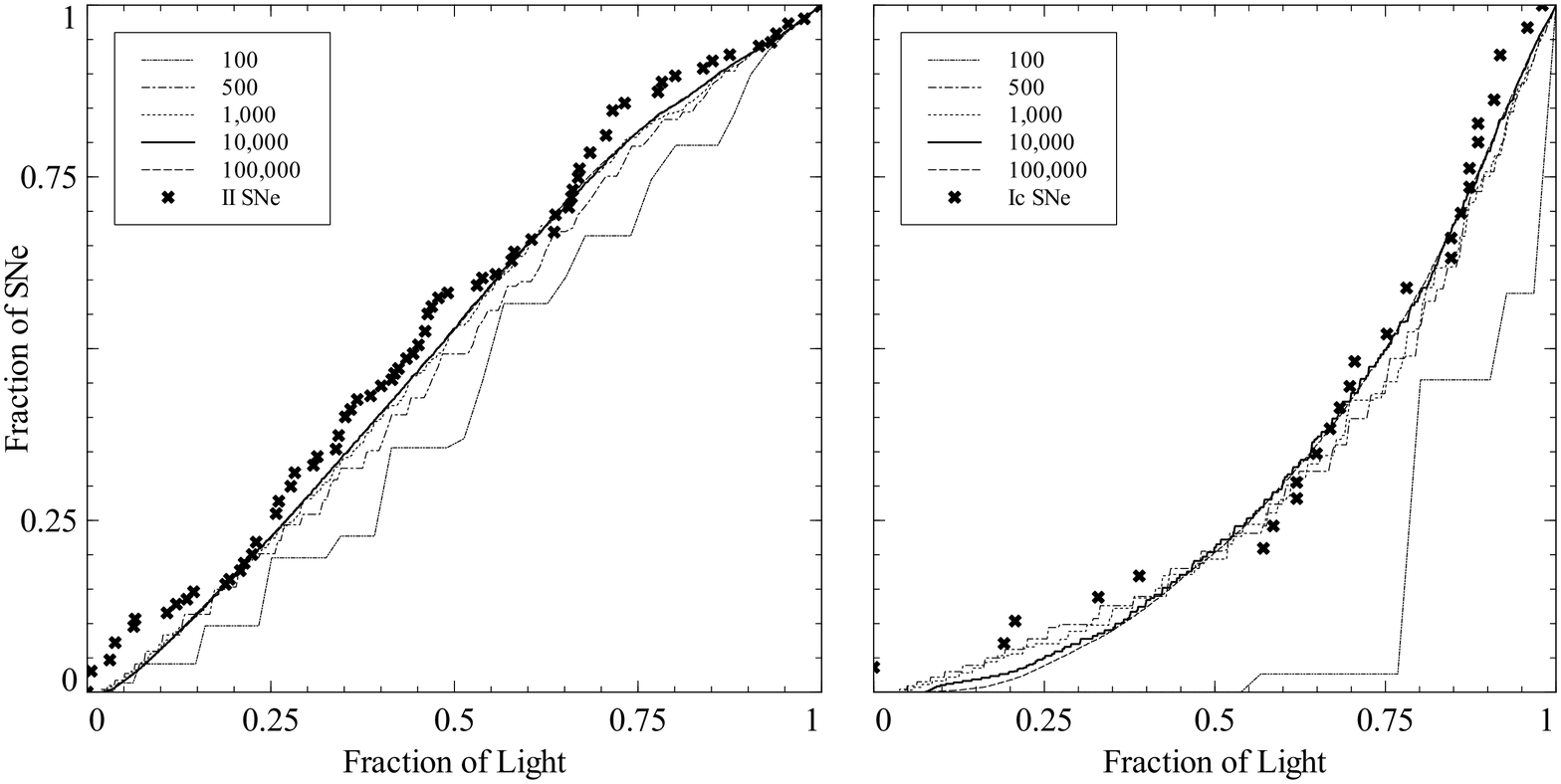}
\caption{Resolution dependence on F06 profiles for $8<M<20$ (left panel) and $M>25$ (right panel), corresponding to type II and Ic SNe progenitors, respectively, as found in \S 2.3. The resolution used throughout this paper, $10^4$ pixels, is indicated by the dark, solid curve.}
\label{fig:resolutions}
\end{figure*}

In the extreme case of too little resolution, one might imagine a host image constrained to a single pixel. In this case the transient location  would not exhibit any dependence on stellar mass, as it would always occur at the same position. To test the effect of image resolution on our results, we compare results for the mass cutoffs found in \S 2.3 at various image resolutions, from 100 pixels, up to $10^5$. In this analysis, absolute pixel resolution and effective resolution are not distinguished so as not to obscure the physical insight with observational complications. Figure \ref{fig:resolutions} demonstrates that the overall nature of the plot does not change once the pixel density goes above a minimum threshold of 500 pixels.
When we repeat the K-S test for the mass cutoff found in \S 2.3 using only 500 pixels as opposed to $10^4$, the confidence level changes by only 2\%. Below the critical threshold of about $10^2$-$10^3$ pixels, however, there simply are not enough data to illustrate a convincing relationship for the locations of transients.

Inclination plays a complicated role within the F06 analysis. At the extreme end, with an edge-on spiral galaxy, the locations of transients can only be said to be near or far from the centers of their hosts, rather than correlated with star-formation. At an intermediate inclination angle, $i$, for a given resolution, the vertical extent of the galaxy contracts by $\cos(i)$, decreasing the total area, and so for each location within the host galaxy, the corresponding fraction of light increases. The global increase in the surface brightness does not play a role in the final F06 analysis, but the decreasing area has the effect of concentrating transients in smaller regions.

\begin{figure*}[ht]
\centering
\includegraphics[width=14cm,height=7.25cm]{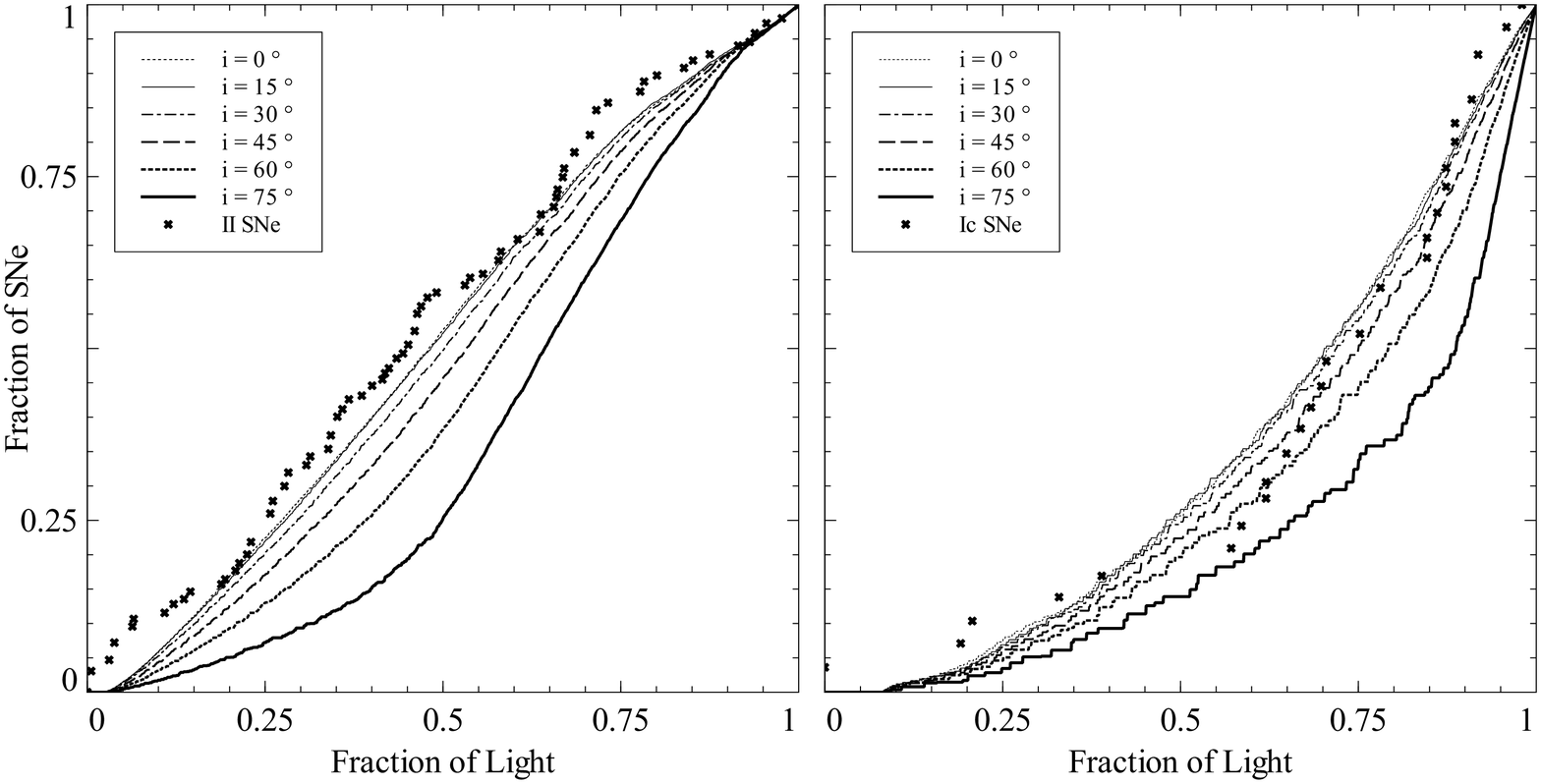}
\caption{Inclination dependence on F06 profiles for $8<M<20$ (left panel) and $M>25$ (right panel), corresponding to type II and Ic SNe progenitors, respectively, as found in \S 2.3.}
\label{fig:inclinations}
\end{figure*}

Figure \ref{fig:inclinations} demonstrates this behavior for the type II and Ic SNe cases.
For $i\leq 45^\circ$, the effect of inclination is marginal, but for larger values of $i$, the F06 profiles rapidly diverge from that of a face-on galaxy. Kelly (2007) specifically excludes ``edge-on" galaxies, placing the upper limit on their sample host inclinations  $i \lesssim 70^\circ$. Excluding any sampling bias, this places the average inclination at $i\approx 35^\circ$, which is below the threshold where such an inclination could be distinguished from face-on in the F06 analysis. Moreover, because high inclinations concentrate transients in brighter regions
for reasons not related to star formation or stellar evolution, galaxies with high inclinations ($i \gtrsim 45^\circ$) are not eligible for the F06 analysis. For example, an F06 analysis comparing observed type II SNe to a simulated host with $i=60^\circ$ would indicate a minimum progenitor mass of $\approx 15$M$_\odot$, which is not consistent with accepted models of stellar physics.

A further complication that inclination could bring is that of the propensity for dust lanes to obscure the brightest regions of spiral galaxies at high inclination. Because we ignore dust and have presented a case for avoiding high inclination spirals, this effect is not treated here.

\section{Gamma Ray Bursts}

Gamma-ray bursts are the brightest observable objects in the universe, far outshining their host galaxies, and they come in two flavors: short-duration bursts and long-duration bursts (eg., Kouveliotou \etal 1993). The favored model for long-duration GRBs, the so-called \textit{collapsar} model, is that of a very massive star whose core collapses to form a black hole (MacFadyen \etal 1999; MacFadyen \etal 2001). As the stellar envelope is consumed by the newly-formed black hole, tremendous energy is released in the form of gamma rays along the axis of rotation, and it is this beaming that gives GRBs their brightness when pointed directly at Earth (Rhoads 1997). The light-curves and spectra of the afterglows of long-duration GRBs occasionally match those of Ic's in their latter stages (Rhoads 1997; Galama \etal 1998; Della Valle \etal 2003; Hjorth \etal 2003; Malesani \etal 2004;  Campana \etal 2006; Pian \etal 2006; Kelly \etal 2007).

Short-duration GRBs, on the other hand, are less luminous than their long-duration counterparts and may or may not be the result of a collimated beaming effect. The model for producing these events has not yet been settled, though possible candidates include neutron star mergers (Eichler \etal 1989; Narayan \etal 1992; Ruffert \& Janka 1999) and magnetar crust failure (Hurley \etal 2005). For the purposes of this paper, only long-duration gamma-ray bursts will be considered and henceforth, simply called GRBs for brevity.

\subsection{Approach}

To accurately model the light distribution of GRB host galaxies, we must construct a reasonable approximation of a low-metallicity, dwarf irregular galaxy that exhibits pseudo-random clustering of star-forming regions. The necessary properties of our model galaxy are:
\begin{itemize}
	\item The average radial brightness must follow an exponential profile (e.g., Barazza \etal 2006).
	\item Starbursts should occur at random intervals for random initial masses, but the time-averaged star formation should yield an exponential profile in total stellar mass (e.g., Barazza \etal 2006).
	\item For simplicity, each starburst in a particular location should be independent of starburst history; that is, the propensity for a starburst to occur in any location should not decrease if that location has already experienced active star formation.
	\item Starbursts should follow the observed initial cluster mass function in their composition (Oey \etal 2003; Dowell \etal 2008). 
	\item Old stellar populations should be diffuse and less concentrated than young populations (Parodi \etal 2003).
\end{itemize}

To build a model that fits these criteria, we first fix the total star formation as a function of radius with an exponential profile
\be
	\dot{\Sigma}(r)=\beta \exp\left(\frac{-r}{r_0}\right),
\label{eq:SigmaStar}
\ee
where $\beta$ is a tunable parameter that sets the total star formation amplitude and $r_0$ is a scale parameter which sets the half-light radius by the relation $R_{hl}\approx 1.68r_0$. Next we fit the initial cluster mass function, which describes how many OB stars, $N_{\rm OB}$, per unit mass are contained within a cluster, as
\ba
\frac{dN}{d\ln N_{\mathrm{OB}}}&\propto&N_{\mathrm{OB}}^{1-\alpha}.
\ea
This can be reproduced by choosing each cluster to contain
\be
	N_{\mathrm{OB}}(\epsilon)=(1-\epsilon)^{\left(\frac{1}{1-\alpha}\right)},
\ee
where $\epsilon$ is a random variable ranging from 0 to 1 and $\alpha = 1.88$ is taken 
to match  the distribution of OB associations in dwarf irregulars 
(Oey \etal 2003; Dowell \etal 2008).

For the last criterion, that old stellar populations will become unbound after $\approx 30$Myr as the gas is ionized and leaves the cluster (Blitz \& Shu 1990; Matzner 2002; Krumholz \etal 2006), resulting in stellar diffusion away from the cluster center, we first calculate the dynamical time for one of these clusters to completely diffuse. Given that the stars in a cluster will have some initial, peculiar velocity, $\Delta v$, we can estimate the diffusion time in the tangential direction as $t_{d,\theta}\approx \left(\pi r/\Delta v\right)$, where $r$ is the radial coordinate of the star cluster within the galaxy. For our model, we use a value for $\Delta v$ on the order of 10km/s (Hunter \etal 2005). The extent to which the cluster has diffused if it is younger than $t_{d,\theta}$ is simply given by $\Delta \theta = (\Delta v/V_{c})[(t_{cl}-30\textrm{Myr})/r]$, where $t_{cl}$ is the age of the cluster and $V_c$ is the circular velocity of the stars at the galaxy's edge.

For radial diffusion, the maximum extent to which the cluster can diffuse can be derived from a simple energy argument given that a dwarf irregular possesses a solid body potential (Binney \& Tremaine 1987) with stellar velocities $v(r)=r\omega$, where $\omega = V_c/R$, with $R$ being the radius of the galaxy and $V_c=v(R)$. If the $r$-coordinate were decomposed into $x$ and $y$ coordinates, $dy/dt = r \omega \cos(\omega t)$, and with a maximal kick in the positive or negative $y$-direction, $(dy/dt)' = (r \omega \pm \Delta v) \cos(\omega t)$ and $y=(r\pm \Delta v/\omega)\sin(\omega t)$. Therefore, $\Delta r_{max} = (\Delta v/\omega)$, regardless of the radial coordinate of the cluster. The diffusion time in the radial direction, then, is $t_{d,r}=\Delta r/\Delta v=1/\omega$ and is the shorter of the two diffusion times, and for our purposes, can be taken to be instantaneous after the cluster becomes unbound. During diffusion, the stars will be arranged according to a Maxwell-Boltzmann distribution with
\ba
	\frac{dn}{d\Delta r}&=&\frac{\Delta r}{\left(R\Delta v/V_c\right)^2}\exp\left[-\left(\frac{\Delta r}{R\Delta v/V_c}\right)^2\right]\label{eq:dndr}\\
	\frac{dn}{d\Delta \theta}&=&\frac{\Delta \theta}{\left(t_\delta\Delta v/r\right)^2}\exp\left[-\left(\frac{\Delta \theta}{t_\delta\Delta v/r}\right)^2\right]\label{eq:dndth},
\ea
where $t_\delta$ here refers to the time since the cluster has become unbound.

\begin{figure*}[ht]
\centering
\includegraphics[width=8.67cm,height=8.67cm]{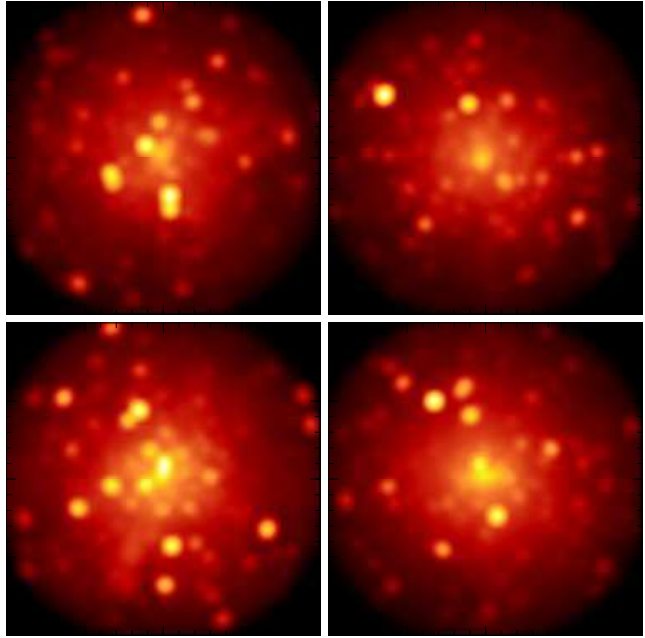}
\caption{Sample dIrr galaxy simulations of B-band luminosity with diameters of 6kpc.}
\label{fig:dIrrs}
\end{figure*}

For each radius, we compile a list of starburst events at random times with a random $\theta$ coordinate until the total number of stars reaches that required by (\ref{eq:SigmaStar}) for that radius over the lifetime of the galaxy. This list is then combined with GALAXEV and equations (\ref{eq:dndr}) and (\ref{eq:dndth}) when rendering the galaxy to incorporate the evolution of the star populations formed in each starburst, the time-variable brightnesses for each OB association, and the dynamical diffusion of each cluster. The result is a model that fits our established criteria for a simplified dIrr galaxy, four examples of which are shown in  Figure \ref{fig:dIrrs}.

\subsection{Tests}

\begin{figure*}[ht]
\centering
\includegraphics[width=8.67cm,height=8.67cm]{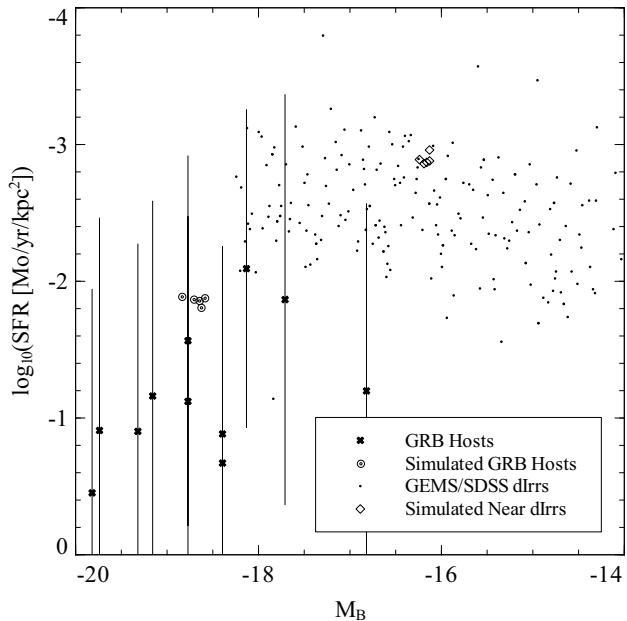}
\caption{Properties of simulated high- and low-redshift dIrrs, as compared to the GRB host dataset from F06, and nearby dIrrs from GEMS and SDSS  (Barazza \etal 2006).}
\label{fig:dIrrGEMS}
\end{figure*}

Before we apply our model to derive F06-style plots, 
we must first ensure that our simulated galaxies match star-formation profiles for observed dIrr GRB hosts, which are generally brighter and have more active star formation than ordinary dIrrs. We accomplish this by adjusting $\beta$ in (\ref{eq:SigmaStar}) to place the resultant log(SFR$/\textrm{kpc}^2$) vs. $\textrm{M}_{B}$ within the GRB host dataset, shown in Figure \ref{fig:dIrrGEMS} with GEMS and SDSS observations plotted for comparison (Barazza \etal 2006; F06). The SFR for the simulated galaxies is a calculation from the resultant galaxy's total stellar mass output over its lifetime rather than the input SFR. The SFR with error bars for the observed GRB hosts are range estimates given the radii and brightnesses of the host galaxies provided by F06, and assuming constant star formation for between 10 Myr and 1 Gyr.

An important test our simulated galaxies must also pass is that of the relation between the scale lengths of the integrated light and of the star formation rate. If the star formation rate remains constant in time, as it does in our model, small stars formed at the edges of the galaxy, where star formation is rare, will persist in the integrated light spectrum for all bands, and this effectively stretches the scale length for the integrated light beyond that of the star formation rate. The effect will obviously be most pronounced in the long-wavelength band-passes, but will also affect the short-wavelength bands as a portion of the output spectrum of old, red stars is also in these bluer bands. Parodi \etal (2003) constrains the relation between the SFR and B-band scale lengths for nearby dIrrs to $r_{0,sfr}\approx (0.86\pm 0.06)r_{0,B}$.  As this is a local constraint it must be compared to dIrrs with somewhat lower SFRs, and thus we need to lower the SFR of galaxies as indicated in Figure \ref{fig:dIrrGEMS}.  Measurements of simulated galaxies with star formation rates and brightnesses that correspond to nearby dIrrs yields $r_{0,sfr}=(0.90\pm 0.07)r_{0,B}$ which is within the error of the Parodi \etal (2003) results.

\subsection{Results}

\begin{figure*}[ht]
\centering
\includegraphics[width=8.67cm,height=8.67cm]{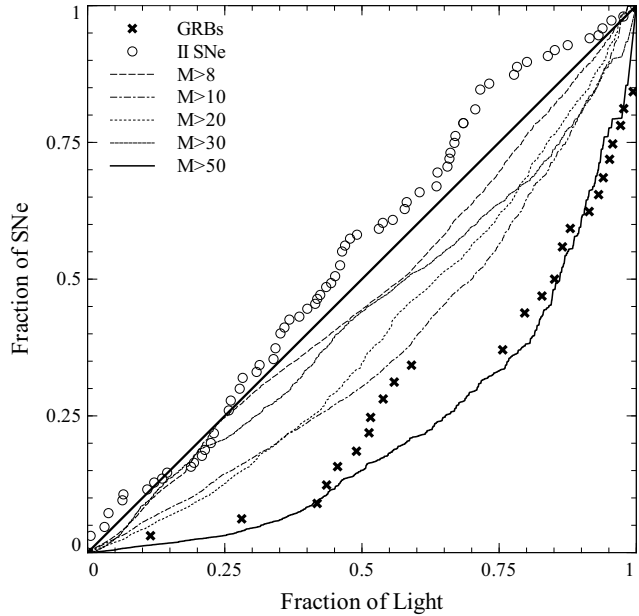}
\caption{F06-style plot for long-duration GRBs in dwarf irregular model. Observed GRBs and type II SNe are plotted for comparison (F06; Kelly \etal 2007).}
\label{fig:fruchtergrbs}
\end{figure*}

\begin{figure*}[ht]
\centering
\includegraphics[width=8.67cm,height=8.67cm]{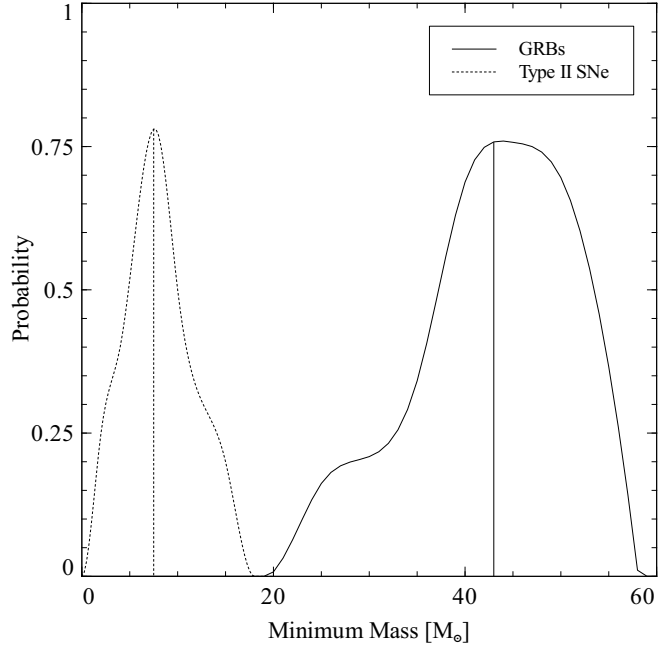}
\caption{K-S test for GRB and type II SNe progenitor masses. A $20\msol$ upper mass cutoff is imposed on the type II simulation as in \S 2, while the GRB simulation has no upper mass cutoff. The observed, most probable minimum mass for GRBs, $\approx 43\msol$, agress with the findings of Larsson \etal (2007).}
\label{fig:ksgrb}
\end{figure*}

As with the type Ic SNe in the spiral model, long-duration GRBs show a strong correlation to high-mass stars when added to  an F06-style plot. Figure \ref{fig:fruchtergrbs} also repeats the earlier observation that type II SNe progenitors are strongly correlated with the B-band light, even in dwarf irregulars. A K-S test, shown in Figure \ref{fig:ksgrb}, reveals that a minimum mass of $\approx 43\msol$ is the most probable candidate for GRB progenitors. This value is in agreement with the findings of Larsson \etal (2007) and supports the conclusion that GRB progenitor stars must be massive enough to form black holes as opposed to white dwarfs or neutron stars (MacFadyen \etal 1999; MacFadyen \etal 2001). One result of particular importance is that though the F06 tracks for GRBs and type Ic's show a characteristic similarity, the masses associated with their progenitors are quite different. This suggests a dependence on the morphological properties of the host galaxy for the F06 method. For the type II SNe probability distribution, again, an upper mass cutoff of $20\msol$ is imposed, yielding a greater probability for $\approx 7.7\msol$ being the minimum mass necessary for a type II SN.

Note that the underlying physics in this comparison is very similar to that in our spiral-wave modeling. In the case of the spiral model, the rotating density wave dynamics was prime motivator for concentrating high-mass, B-band emitters in the brightest parts of the galaxy. In the dwarf irregular case, the dynamics of stellar clusters play this role, concentrating young, blue stars nearest the locations of their birth and hence near the brightest portions of their host galaxies. 

\section{Type Ia SNe}

Lastly, we apply our approach to type Ia SNe, the so-called \textit{standard-candle} events by which distances to host galaxies can be measured (Colgate 1979; Branch \& Tammann 1992; Phillips 1993). These are thought to be the result of a white-dwarf accreting mass above the Chandrasekhar limit and collapsing into a neutron star either via stable accretion from a binary (Nomoto 1982; Starrfield 1994; Webbink 1994) or a collision with another white dwarf (Woosley \& Weaver 1994).  Like type Ic SNe, type Ia's are often found in late-type, spiral galaxies, and we focus purely on the populations found in such hosts. In modeling the locations of these transient events, we must take into account the different processes that lead to type Ia's as compared to core-collapse SNe. Specifically, the incidence of type Ia's is not directly related to progenitor mass.

\subsection{Approach}

Type Ia's have been observed to have two components: a larger component that is dependent on the star formation rate and the time since star-formation, and the smaller component that is SFR-independent(Mannucci \etal 2005; Scannapieco \& Bildsten 2005):
\be
	\frac{SNR_{\mathrm{Ia}}(t)}{(100\textrm{yr})^{-1}} =
	A\left[\frac{M_*(t)}{M_{\sol}}\right]
	+B\left[\frac{\dot{M}_*(t-\tau)}{M_{\sol} \textrm{yr}^{-1}}\right],
\ee
where $M_*$ is the total stellar mass and $\dot{M}_*(t-\tau)$ is the star formation rate at some previous time, $t-\tau$. Scannapieco \& Bildsten (2005) presented the formalism for constraining the values of A and B, and Sullivan \etal (2006)  find $5.3\pm 1.1 \times 10^{-14}$ and $3.9\pm 0.7 \times 10^{-4}$, respectively. 

This relation establishes all the necessary information for populating Ia events in our model galaxy without making any assumptions about progenitor masses. Whereas in the core-collapse model, stellar evolution and lifetimes were the dominant factors in placing the transient events in particular spatial locations, in the Ia model, the locations of the B component transients are controlled by a characteristic delay-time, $\tau$, (\eg Madau \etal 1998; Gal-Yam \& Maoz 2004) in the perturbation after star formation. 

\begin{figure*}[ht]
\centering
\includegraphics[width=8.67cm,height=8.67cm]{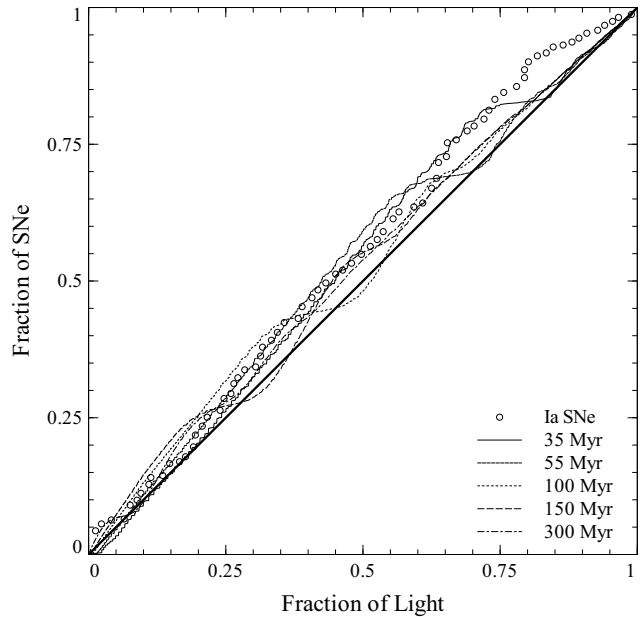}
\caption{F06-style plot for Ia SNe in spiral model with several delay times. Observed type Ia SNe are plotted for comparison (Kelley \etal 2007).}
\label{fig:FruchterIaBad}
\end{figure*}

When plotting the Ia distributions in an F06 fashion for the B-band, the effect of changing the delay time for the B component is not immediately noticeable, as shown in Figure \ref{fig:FruchterIaBad}. To better constrain the delay time of the B component, we suggest a modification to the F06 method. In this case, the image of each host galaxy is normalized by its overall radial, exponential profile, which emphasizes the spiral density wave distribution with respect to the overall stellar distribution. That is for each galaxy we take
\be
\tilde \Sigma_\nu(r,\theta,t) \equiv  \Sigma_\nu(r,\theta,t)/\phi(r),
\ee
where $\phi(r)$ is the average radial profile of the galaxy. Similarly, each of the type Ia transient event images is normalized by this same exponential profile, with
\be
\tilde \Sigma_{\rm SN}(r,\theta,t) \equiv  \Sigma_{\rm SN}(r,\theta,t)/\phi(r).
\ee

Each of these normalized distributions is then used to calculate the normalized light in pixels fainter than each galaxy, that is with central pixels artificially dimmed. Finally, these distributions are combined into F06-style plot using the normalized number of Ia's, that is with Ia's occurring near the outskirts of disk galaxies, contributing more than those near the crowded centers of these galaxies. This simple procedure draws out the spiral-wave structure, resulting in large differences between different delay-time distributions. 
\begin{figure*}[ht]
\centering
\includegraphics[width=8.67cm,height=8.67cm]{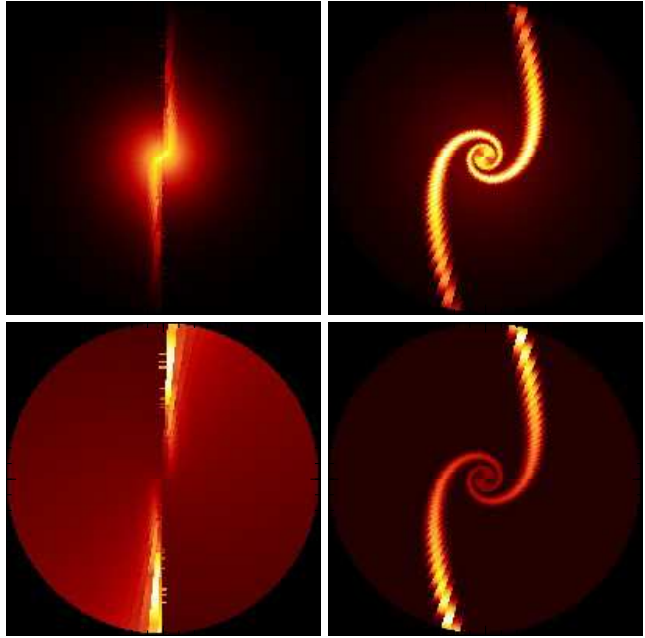}
\caption{A B-band luminosity profile for a spiral galaxy and a SNe profile for $\tau = 35$ Myr, respectively. The bottom pair have been normalized by an exponential to remove the non-dynamical component of the Ia SNR.}
\label{fig:IaSpirals}
\end{figure*}

Because this method relies on transient locations being dependent on the ordered dynamical properties of the host in conjunction with the delay time, it is most useful for applying to spiral galaxies rather than irregulars. Moreover, since ellipticals do not experience any star formation, they are also not eligible for this analysis, or else $\tau$ would have to be inordinately large.

\subsection{Results}

\begin{figure*}[ht]
\centering
\includegraphics[width=8.67cm,height=8.67cm]{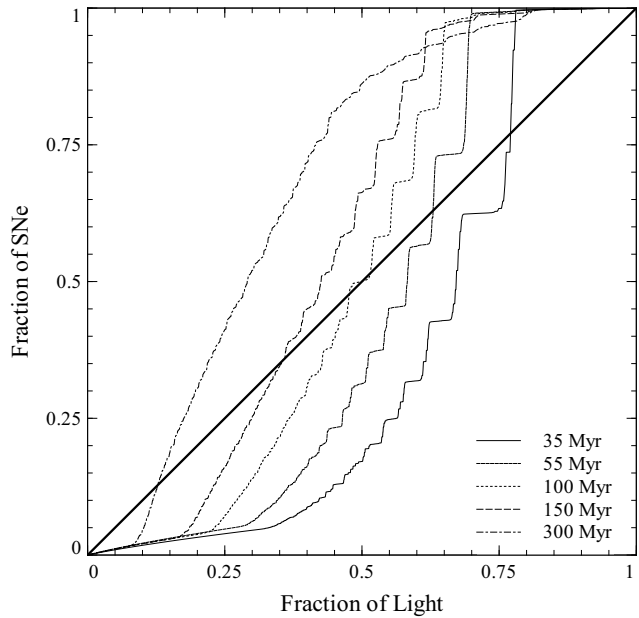}
\caption{Modified F06-style plot for Ia SNe in spiral model, in which each pixel is normalized by $\phi(r)$.}
\label{fig:FruchterIa}
\end{figure*}

In Figure \ref{fig:FruchterIa}, we show the results of carrying out this modified F06 analysis on our model disk galaxies. Given that regardless of the candidate progenitor model for type Ia's (binary accretion, white dwarf mergers, etc.), a WD star must have first been produced, the minimum value for $\tau$ is necessarily the shortest lifetime of a white dwarf progenitor star. Core-collapse occurs for stars of mass greater than $8M_{\sol}$ which corresponds to a lifetime of approximately 35 Myr at solar metallicity, and so $\tau \geq 35$ Myr. This minimum value is about 10$\%$ of the dynamical time of the model galaxy, and thus we may vary $\tau$ by as much as an order of magnitude while still retaining the usefulness of the F06 analysis.

What is evident from the modified F06 plot is that with an increasing delay in the onset of the B component, the Ia profile crosses the integrated light curve at increasingly dimmer fractions of the total light. Figure \ref{fig:IaSpirals} demonstrates why this is so. As the delay time for the dynamical component increases, the peak output of the SNe intensity shifts to angular coordinates  correspond to ever decreasing luminosities in the integrated light, causing the crossover in the modified F06 plot at lower fractions of total light.

This behavior, when matched against observations that employ our modified analysis, should serve well to constrain the time delay between star formation and Ia transient events, thus placing limits on what kinds of progenitor models are viable explanations for type Ia SNe. The accretion model requires only that one white dwarf has formed in binary with an evolved star and would be consistent with a value of $\tau \approx 35$ Myr, while the white dwarf merger model introduces a degeneracy and would likely push the value of $\tau$ to longer time-scales. Furthermore, this modified procedure can be simply applied to existing data sets, an application which we are actively pursuing.

\section{Conclusions}

The F06 method is a valuable tool in the search for transient progenitors. By relying on simple observations, the method remains agnostic to the details of the transient events, such as spectroscopic signature or peak intensity, and instead places concise and testable constraints on their spatial distributions. 

In this paper, we have demonstrated first, that models of spiral galaxies illustrate the dynamics that lead to the observed spatial distributions of core-collapse SNe. In a spiral galaxy, the rotating star-formation wave is the primary dynamical component for setting the spatial distributions of SNe with varying progenitor masses. The distances that stars are able to travel from this wave during their lifetime determines how F06 profiles for certain transient events will present themselves. Since there is a direct relationship between stellar lifetime and mass, our analysis has constrained type Ic SNe only to stars above $\approx 25\msol$ and has supported estimates for the minimum mass of type II SNe of $\approx 8\msol$.

In the dwarf irregular model, the role of the star-forming wave is played by the star-formation profile which, in conjunction with the cluster diffusion, places young stars in the brightest portions of the galaxy. By employing GALAXEV results in combination with observed dependencies on SFR and the IMF, our dwarf irregular model fulfills our initial criteria for being a representative counterpart of real galaxies as well as adhering to the observed integrated light profiles of dwarf irregulars. In combining the F06 method with the results of this model, we are able to place constraints on the progenitor masses of GRBs to stars with masses above $\approx 43\msol$. Our dIrr models also confirm a minimum progenitor mass for type II SNe of $\approx 7.7\msol$.  In future work, we will improve our simulation by adding a reasonable amount of patchy absorption in the dwarf and spiral galaxies. With too much dust, the GRB results would not conform to observational constraints, and so, these models might also serve to constrain the effects of dust in GRB hosts.

For type Ia SNe, the delay time associated with the dynamical component of the two-channel Ia intensity profile is the primary factor in determining their spatial distribution. However, the effect of changing the delay time for the dynamical component is not evident in a standard F06 plot. For this reason, we have suggested a modification to this analysis that normalizes each pixel by overall radial profile of the host.  Applying this method to our simulated galaxies  demonstrates a clear dependence of the resultant profile on the delay time of the dynamical component. Future application of this method to existing observations of type Ia SNe will serve to place concise constraints on the delay time, supporting or refuting candidate progenitor models for SNe Ia. 

For each of the objects we have considered in this paper, we have demonstrated that galactic dynamics play an important role in the distributions of stars at the moment of their deaths.  Furthermore, while not considered here, this techniques is likely to be applicable to other types of objects such as classical novae, Wolf-Rayet stars, and planetary nebulae.  From the observed distributions of transients, we are able to constrain the masses and properties of the progenitors from which they are formed. As transients flicker in and out of existence, the physics of stars is written across the night sky.

\acknowledgments

We are grateful to Lars Bildsten, Bruce Doak, Brian Gleim, and Sangeeta Malhotra for their many useful comments and suggestions. We are also grateful to the referee for their many useful comments and suggestions.

\end{document}